\documentclass{pasj00}
\draft
\usepackage{natbib}

\begin{document}
\SetRunningHead{Isobe et al.}{Hinode white light flare}
\Received{2000/12/31}
\Accepted{2001/01/01}

\title{Flare Ribbons Observed with G-band and Fe \emissiontype{I} 6302\AA \, Filters of 
the Solar Optical Telescope on Board Hinode}



%
\author{
   Hiroaki \textsc{Isobe}\altaffilmark{1}, 
   Masahito \textsc{Kubo}\altaffilmark{2,3}, 
   Takashi \textsc{Minoshima}\altaffilmark{1}, 
   Kiyoshi \textsc{Ichimoto}\altaffilmark{4}, 
   Yukio \textsc{Katsukawa}\altaffilmark{4}, 
   Theodore D. \textsc{Tarbell}\altaffilmark{5}, 
   Saku \textsc{Tsuneta}\altaffilmark{4}, 
   Thomas E. \textsc{Berger}\altaffilmark{5}, 
   Bruce W.  \textsc{Lites}\altaffilmark{2}, 
   Shin'ichi \textsc{Nagata}\altaffilmark{6}, 
   Toshifumi \textsc{Shimizu}\altaffilmark{3}, 
   Richard A. \textsc{Shine}\altaffilmark{5}, 
   Yoshinori \textsc{Suematsu}\altaffilmark{4}, 
   and
   Alan M. \textsc{Title}\altaffilmark{5}}
 \altaffiltext{1}{Department of Earth and Planetary Science, University of Tokyo, 
   Hongo, Bunkyo-ku, Tokyo 113-0033}
 \email{isobe@eps.s.u-tokyo.ac.jp}
 \altaffiltext{2}{High Altitude Observatory, National Center for Atmospheric Research, P. O. Box 3000, 
   Boulder, CO 80307, USA}
 \altaffiltext{3}{Institute of Space and Astronautical Science, Japan Aerospace Exploration Agency, 
    3-1-1 Yoshinodai, Sagamihara, Kanagawa 229-8510}
 \altaffiltext{4}{National Astronomical Observatory of Japan, 2-21-1 Osawa, Mitaka, Tokyo 181-8588}
 \altaffiltext{5}{Lockheed Martin Advanced Technology Center, O/ADBS, B/252 3251 Hanover Street, 
    Palo Alto, CA 94304, USA}
 \altaffiltext{6}{Hida Observatory, Kyoto University, Kamitakara, Gifu 506-1314}


\maketitle

\begin{abstract}
The Solar Optical Telescope (SOT) on board Hinode satellite observed an X3.4 class flare on 2006 
December 13. Typical two-ribbon structure was observed, not only in the chromospheric Ca\emissiontype{II} H line 
but also in G-band and Fe\emissiontype{I} 6302 \AA \, line. The high-resolution, seeing-free images achieved 
by SOT revealed, for the first time, the sub-arcsec fine structures of the ``white light" flare. 
The G-band flare ribbons on sunspot umbrae showed a sharp leading edge followed by a diffuse inside, 
as well as previously known core-halo structure. The underlying structures 
such as umbral dots, penumbral filaments and granules were visible in the flare ribbons. 
Assuming that the sharp leading edge was directly heated by particle beam and the diffuse parts were 
heated by radiative back-warming, we estimate the depth of the diffuse flare emission using the 
intensity profile of the flare ribbon. We found that the depth of the diffuse emission is about 100 km 
or less from the height of the source of radiative back-warming. 
The flare ribbons were also visible in the Stokes-V images of Fe\emissiontype{I} 6302 \AA \, as a transient 
polarity reversal. This is probably related to "magnetic transient" reported in the literature. 
The intensity increase in Stokes-I images indicates that the Fe\emissiontype{I} 6302 \AA \, line was 
significantly deformed by the flare, which may cause such a magnetic transient.
\end{abstract}

\section{Introduction}
Solar flares that are visible in optical continuum are called white light flares (WLFs). 
They have been of interest because the radiative loss by the WL emission is 
significant, they are similar to stellar flares in many aspects, and their energy transport 
and emission mechanisms have been challenging problems \citep{neidig89}.  
Observers used to believe that WL emissions are seen only in the most energetic flares 
\citep{hiei82, neidig89}. 
However, seeing-free observation from spacecrafts have revealed that even C-class 
flares show enhancement in optical continuum \citep{matthews03, hudson06}. 

Close correlation in space and time between WL and hard X-ray (HXR)  emissions have 
been reported in many events, strongly suggesting that WLFs are energized by nonthermal 
electron beams \citep{rust75, hudson92, neidigkane93, chending06}. 
Recently, several authors studied the energetics of WLFs, examining 
whether nonthermal electron beam had enough energy to power the WLFs 
\citep{metcalf03, chending06, fletcher07}.  
Their results indicate that, for most of WLFs, the energy required to power the WL emission 
was comparable to the total energy carried by the electron beam, and therefore the WLFs were 
energized by the low energy electrons,  about 20 keV or less, which carried most of the energy.
Such low energy electrons cannot penetrate into the photosphere or lower chromosphere,  
so they deposit the energy in the upper chromosphere \citep{hudson72}. Hence, the WL emission may also 
come from the upper chromosphere, or the energy is transported by radiation 
into deeper layer at which the WL emission occurs (so called back-warming effect; 
\citealt{machado89, metcalf90}). 

Morphology also gives some clues to examine the mechanisms of WLFs. 
High-resolution images  often show separation of a  WLF kernel into a bright inner core 
and a fainter outer region (halo) \citep{neidig93, hudson06}. 
Using high-cadence ground based observation, \citet{xu06} examined the timing of WL 
and HXR emission in the core and halo separately. They found that the peak of WL in the halo 
showed a delay from that of HXR, but they were almost simultaneous in the core. This suggests 
direct heating by electron beams in the core and indirect heating  by back-warming 
(or something else) in the halo. 

The high-resolution, seeing-free images from the Solar Optical Telescope (SOT; Tsuneta et al. 2007) 
on board Hinode (Kosugi et al. 2007) provide a qualitatively new data set for studying WLFs. 
After its launch in 2006 September, Hinode has successfully observed several flares. 
The X3.4 flare on 2006 December 13 was of particular interest from the view point of WLF, 
because the SOT observed clear two ribbons not only in the chromospheric Ca\emissiontype{II} H images 
but also in the G-band images and  the Stokes I and V images of Fe\emissiontype{I} 6302 \AA \, line. 
G-band and the Fe\emissiontype{I} 6302 line are not ``white light" in a strict sense, but they both originate  
in the photosphere. Moreover, the flare emission seen in Stokes V images may be related 
to so called magnetic transients (magnetic anomalies), the transient signals in magnetograms  
associated with WLFs, whose origin remains uncertain \citep{patterson81, zirin81, kosovichev01, 
qiu03}. 
The aim of this paper is to present the initial SOT observation of the flare ribbons of 
this event and discuss their possible indications for the mechanisms of WLFs.

\section{Observation}
\subsection{Data}
The X3.4 class flare on 2006 December 13 in active region 
NOAA10930 was observed by 
all three instruments on board  Hinode: 
the SOT, 
the Extreme-ultraviolet Imaging Spectrometer (EIS; Culhane et al. 2007), 
and the X-Ray Telescope (XRT; Golub et al. 2007). 
Kubo et al. (2007) studied the long-term evolution of vector magnetic field of this region using 
the SOT data. 

The aperture of SOT is 50 cm and is the largest for an optical telescope ever build to observe 
the Sun from space. The initial investigation has shown that the SOT achieved its diffraction 
limit, 0.2-0.3 arcsec.  
The SOT performs both filter and spectral observations at high polarimetric precision 
(Ichimoto et al. 2007). 
The Spectro-Polarimeter (SP) obtains the full Stokes profiles of the magnetic sensitive 
Fe\emissiontype{I} 6302\AA \, line. 
The Filtergram (FG) observation has two channels, namely the Broadband Filter Imager (BFI) 
and the Narrowband Filter Imager (NFI). 
BFI has 6 broad bands (CN band, Ca\emissiontype{II} H line, G-band, and 3 continuum bands) and 
obtains photometric images with 0.0542 arcsec/pixel sampling. 
NFI provides intensity, Doppler, and Stokes polarimetric imaging with 0.08 arcsec/pixel sampling. 
There are 10 available spectral lines including Fe lines with different magnetic sensitivity, 
Mg\emissiontype{I}b, Na D, and H$\alpha$. 
For instrumental details of the SOT, see Tsuneta et al. (2007), Suematsu et al. (2007), 
Ichimoto et al. (2007), and Shimizu et al. (2007). 

The SOT/FG observation has a large degree of freedom: choice of filters, 
field of view (FOV), pixel summing, cadence etc.  
During the flare, BFI obtained Ca H and G band images with full field of view 
\timeform{218"}$\times$\timeform{109"}, 
2$\times$2 summing and 2 min cadence. No continuum band images was taken during this flare.   
NFI obtained the intensity (Stokes-I) and circular polarization (Stokes-V) 
at -120 m\AA \, offset of the photospheric line Fe\emissiontype{I} 6302. Hereafter we refer  
them as 6302I and 6302V images, respectively. 
The FOV, pixel summing and cadence of NFI were \timeform{328"}$\times$\timeform{164"}, 
2$\times$2, and 2 min, respectively. 
Eight exposures during the polarization modulation were used to 
obtain a set of 6302I and 6302V images, which took about 14 sec. 
This results in some blurring of the image, particularly when sharp structures are 
moving rapidly, as is the case of the flare ribbons. 

The 6302V images provide a proxy for line-of-sight magnetograms. 
The line-of-sight magnetic field $B_l$ is approximately given by 
\begin{equation}
B_l=-\frac{C_V}{0.798C_I}\times 10000 \; {\rm (G)}, 
\end{equation}
where $C_I$ and $C_V$ are the counts of the 6302I and 
6302I images, respectively, and $C_V/0.798$ gives 
the circular polarization (Stokes-V). This approximation is crude and  
breaks down for large $B_l$. For example, $C_V$ in the umbrae is smaller than that 
in the penumbrae.   

\begin{figure}
  \begin{center}
    \FigureFile(160mm,160mm){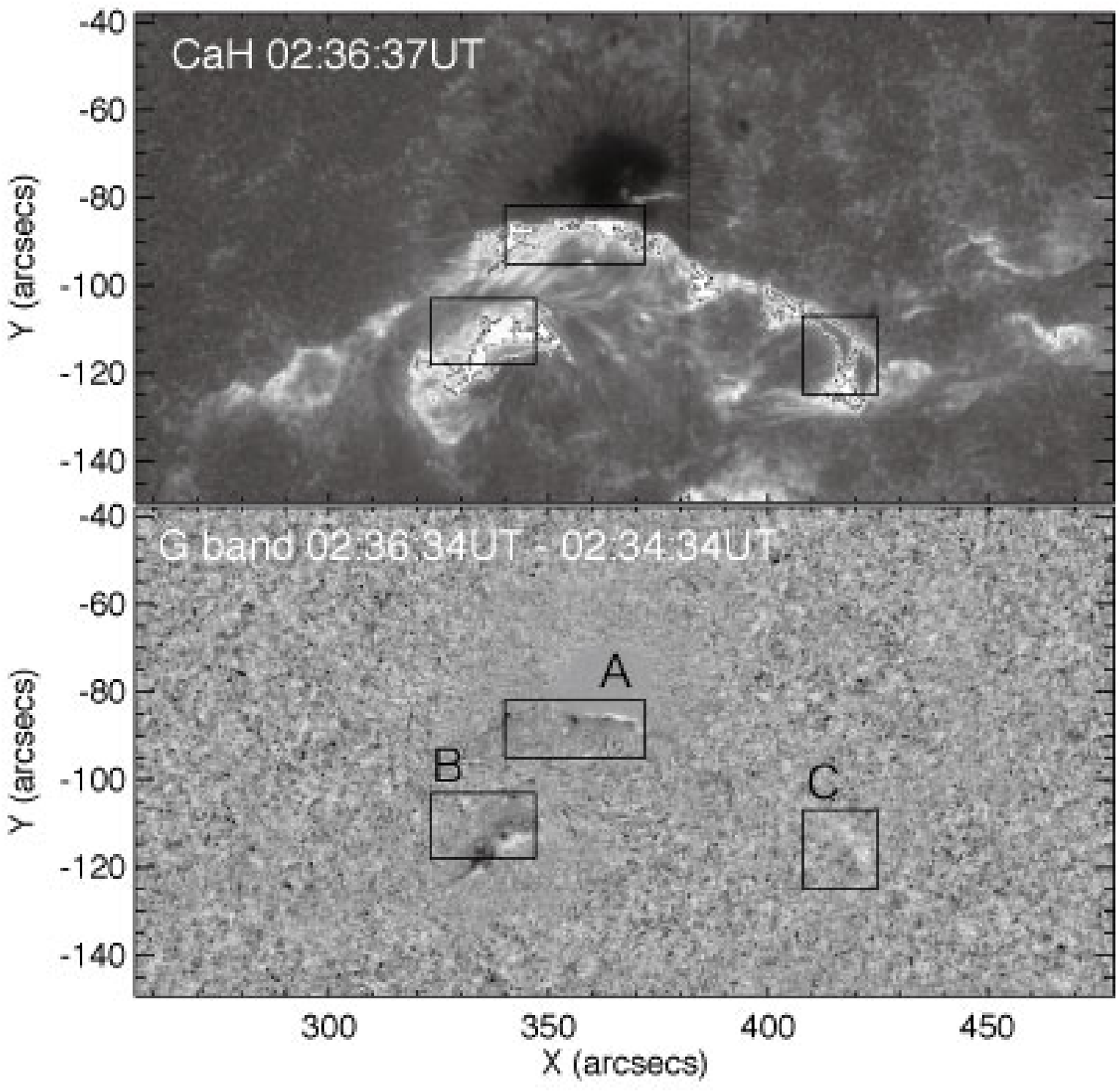}
  \end{center}
  \caption{Upper panel: Ca H image of the X3.4 flare on 2006 December 13. 
  Lower panel: running difference of the G-band images. 
  }\label{fig:fig1}
\end{figure}

 Figure \ref{fig:fig1} shows a Ca H image and a running difference 
 of G-band images. 
 The flare occurred between two umbrae which have opposite polarities, where strong 
shearing motion was found (Kubo et al. 2007). 
The Ca H image shows large two ribbons that extend  nearly 200 arcsec along 
the polarity inversion line. Unfortunately the exposure was not optimized 
for a flare and hence the brightest parts of the Ca H ribbons are saturated. 
The G-band images also shows clear ribbon structures, but they were visible 
only some parts of the CaH ribbon. Indeed the G-band ribbons are co-spatial 
with the saturated, i.e. the brightest, parts of the Ca H ribbon, as indicated 
by the box A (upper sunspot), box B (lower sunspot) and box C (outside 
sunspots) in figure \ref{fig:fig1}. The flare ribbons are also visible 
in 6302 I and 6302V with less spatial extent. 
In the following we examine the fine structures in the different parts of the ribbons.

\subsection{Ribbon in the Northern Sunspot} 
\begin{figure}
  \begin{center}
    \FigureFile(160mm,80mm){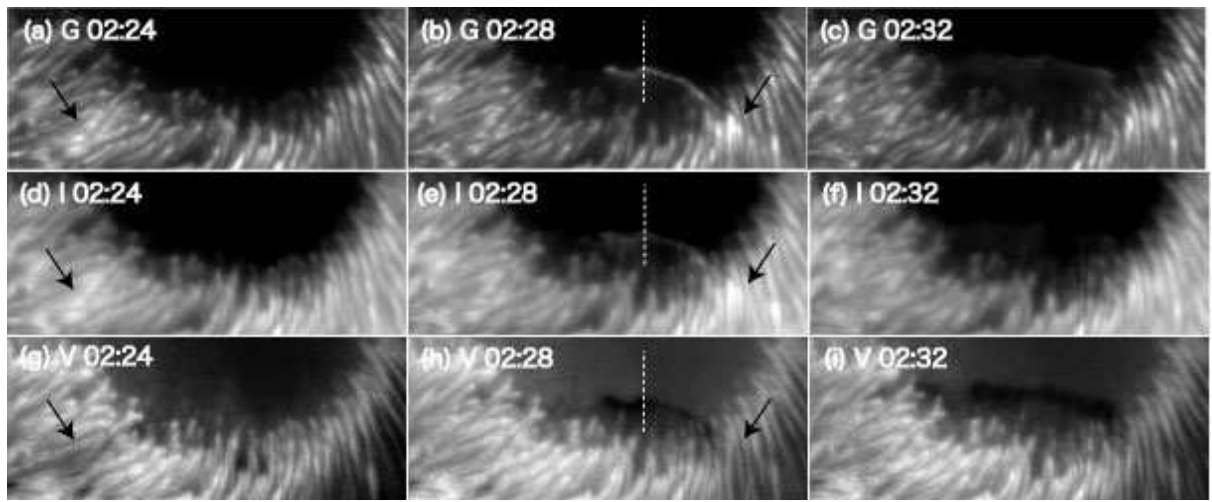}
  \end{center}
  \caption{Snapshots of flare ribbons in the northern sunspots in G-band (panels a-c), 
  6302I (panels d-f), and 6302V (panels g-i). The FOV is the same as box A in figure \ref{fig:fig1}, 
 and  the times of the images are indicated in the figure. The arrows indicate the 
  flare ribbon in the penumbra. The intensity profiles along the dashed vertical lines 
  are plotted in figure \ref{fig:fig3}. 
  A movie is provided as an on-line material (movie 1).   
  }\label{fig:fig2}
\end{figure}

Figure \ref{fig:fig2} shows G-band, 6302I, and 
6302V images of the ribbon in the northern sunspot. The time of 
the images are 02:24, 02:28 and 02:32, and the FOV is the same as box A in figure \ref{fig:fig1} 
(see also movie 1). 
The remarkable features is the sharp leading edge followed by the diffuse inside. 
The leading edge and the diffuse inside look similar to the core-halo 
structure found in previous observations \citep{neidig93, hudson06}, 
but such a sharp and coherent leading edge has not seen before. 
The leading edge is the sharpest in G-band, and also visible in 6302I, but not clear in 6302V. 

The intensity profiles along the vertical dotted line in figure \ref{fig:fig2} are shown in 
figure \ref{fig:fig3}. The diamonds are those at 02:28, and the plus signs are those at 
02:20.  The peak intensity on the leading edge is 600\% larger than the 
pre-flare intensity in G-band, and 50\% larger in 6302I. The 6302V count reverses the sign, 
from $\sim$ 100 in the pre-flare umbra to $\sim$ -170 at the peak. 
The sign reversal is transient and occurs only when and where the flare ribbon was 
also visible in 6302I. Therefore it is very unlikely a real magnetic 
field change. 

The flare ribbon is also visible in the penumbra, as indicated by arrows in figure \ref{fig:fig2}. 
By comparison of the images at different time, one can recognize the same filamentary structure of the 
penumbra in the flare ribbon, both in G-band and in 6302I. 
Thin gives us an impression that the flare component seen in these filters are optically thin. 
The intensity increase in penumbral flare ribbon is $\sim$ 50\% for G-band and 
$\sim$ 10\% for 6302I. The 6302V images also shows the flare ribbon in the 
penumbra.  It does not change the sign but decreases significantly in magnitude, 
typically from 500 to 300. 

\subsection{Ribbon in the Southern Sunspot}
Figure \ref{fig:fig4} shows the G-band images of the flare ribbon on the 
southern sunspot (see also movie 2). The FOV is the same as box B in figure \ref{fig:fig1}. The leading-edge 
and diffuse-inside structure is also seen, but there are more patchy bright points compared with the 
ribbon in the northern sunspot. Note that in this sunspot there are many umbral dots, 
which are almost absent in the northern sunspot. 
Many of the bright points in the flare ribbon seen at 02:28UT are co-spatial with the 
umbral dot seen in before (02:24) and after (02:32) the ribbon passage. One of the 
example is indicated as ``UD" in figure \ref{fig:fig4}. Hence these bright 
points are not the intrinsic structure of the flare emission. 

However, some bright points are not the underlying umbral dots. An example is 
indicated as ``FK" in figure \ref{fig:fig4}. The intensity profiles across the umbral dot ``UD" and 
the flare kernel ``FK" (along the white lines with a gap in the middle) are shown in panel (d) of 
figure \ref{fig:fig4}. Contrary to UD, FK shows no enhancement before and after the 
ribbon passage. Therefore these flare kernels are not due to the pre-flare atmospheric 
structure but indeed due to the fine structure in the flare energy deposition. 
The FWHM of the FK is about 300 km, but we cannot rule out the existence 
of unresolved finer structure. 

The 6302I images (not shown) were similar to those in G-band, showing umbral dots, flare kernels 
and diffuse ribbon, although the spatial resolution is less and the images look more blurred. 
In 6302V images, the diffuse ribbon was seen but the flare kernels were unclear. It 
also showed sign reversal where the flare ribbon is bright in 6302I.  

\subsection{Ribbon outside the Sunspots}
The flare ribbon is also visible outside the sunspot, as 
shown in figure \ref{fig:fig5}. The FOV is the same as box C in figure \ref{fig:fig1}. 
Panel (a) and (b) shows the successive G-band images in which the flare ribbon is seen. 
The intensity increase in the flare ribbon is 10-20\%, which is comparable to the contrast 
of background granulation. The flare ribbon is more clearly seen in the differenced image (panel c). 
It should be noted that the underlying granules are visible in the flare ribbon. 
The flare ribbon looks diffuse, but the existence of fine structures is difficult to 
address because of the underlying granules. 

Panel (d) of figure \ref{fig:fig5} shows the difference image of 6302I at almost 
the same time as the G-band difference image. The intensity increase is about 
5\%. We could not recognize the flare ribbon in 6302V images in this region.

\section{Discussion}

\subsection{Height of the Flare Emission}

\begin{figure}
  \begin{center}
    \FigureFile(140mm,60mm){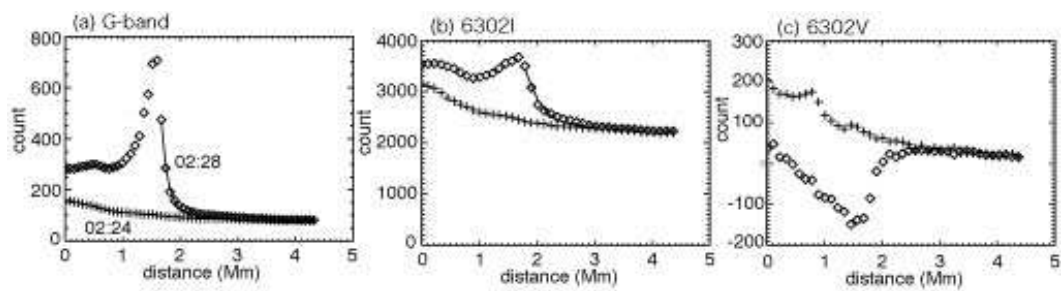}
  \end{center}
  \caption{Intensity profiles along the vertical line in figure \ref{fig:fig2}. 
  Plus signs and diamonds correspond to 02:24UT (preflare) 
  and 02:28UT (during flare), respectively. The solid lines shows the best-fit model; 
   see section 3.1.  
  }\label{fig:fig3}
\end{figure}

\begin{figure}
  \begin{center}
    \FigureFile(140mm,100mm){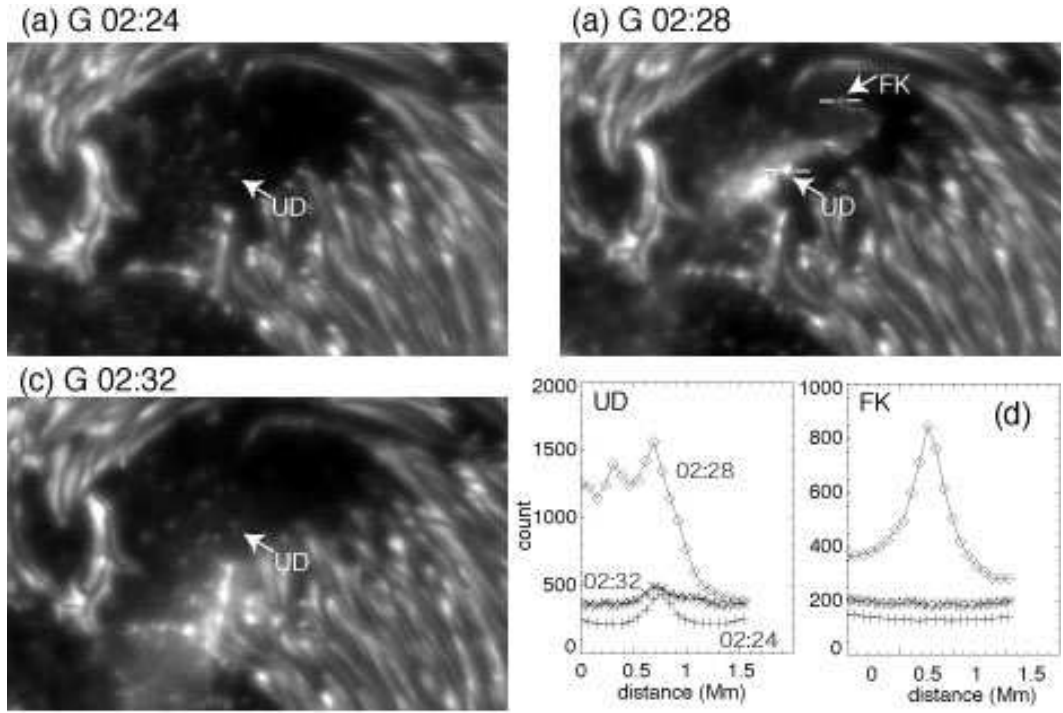}
  \end{center}
  \caption{(a)-(c): G-band images of the flare ribbon in the southern sunspot. 
  FOV is the same as box B in figure \ref{fig:fig1}. (d): Intensity profiles of the 
  umbral dot (marked as UD) and the flare kernel (marked as FK). 
  A movie is prvided as an on-line material (movie 2). 
  }\label{fig:fig4}
\end{figure}

\begin{figure}
  \begin{center}
    \FigureFile(60mm,60mm){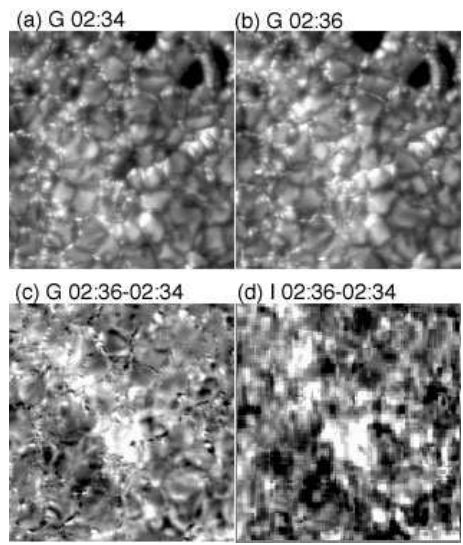}
  \end{center}
  \caption{(a)-(b): G-band images of the flare ribbon outside the sunspots. 
  FOV is the same as box C in figure \ref{fig:fig1}. (c) Difference image of G-band. 
 (c) Difference image of 6302I.   
  }\label{fig:fig5}
\end{figure}

The height of the WLF emission is closely related to the transport and emission mechanisms.  
The striking sharpness of the fine structures in the flare ribbons found by the SOT allows us to 
put a constraint on their vertical structure, as discussed below. 
It seems reasonable to believe that the sharp leading edge and diffuse inside found 
in the present event may be the analog of the core-halo structure found in previous observations 
\citep{neidig93, hudson06, xu06}. Assuming that the interpretation 
of \citet{xu06} can be applied to our event, i.e., direct heating by particle beams in the 
core (leading edge) and radiative back-warming in the halo (diffuse part), 
we made a simple model to measure 
the upper limit of the depth of the layer at which the diffuse emission occurs. 
The intensity profiles plotted in figure \ref{fig:fig3} show that there is a 
tail outside the leading edge. We assume that the flare energy is deposited 
only at the peak of the leading edge, and subsequent back-warming by the 
irradiation from the peak produces the outside tail. Using the intensity profile of the 
outside tail,  we can derive the depth of the diffuse emission layer relative to 
the irradiating source. To do this, we make several more assumptions: 
(1) Variation along the flare ribbon is neglected, i.e., the system is two-dimensional, 
(2) The pre-flare atmosphere is uniform, 
(3) The irradiation is isotropical.  
(4) The source of irradiation is a one-dimensional line whose thickness is negligibly small. 
Thus we neglect the irradiation from the inner part of the ribbon. 
(5) The diffuse emission from the outside tail comes from an infinitely thin layer 
with fixed geometrical height, at which the irradiating photons are absorbed. 
(6) Absorption of the irradiating photons between this layer and the irradiating source 
is negligible. 

\begin{figure}
  \begin{center}
    \FigureFile(60mm,60mm){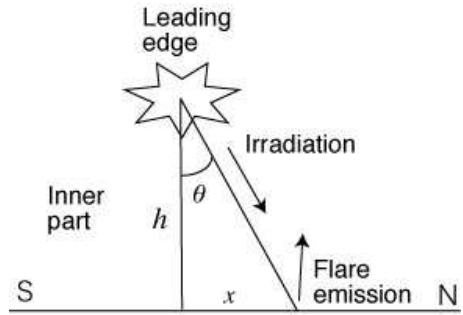}
  \end{center}
  \caption{Schematic illustration of the model. 
  }\label{fig:fig6}
\end{figure}

The model is illustrated in figure \ref{fig:fig6}. Here $h$ denotes the depth of 
the flaring layer by back-warming, measured from the irradiation source. $x$ is 
the distance from the irradiation source projected on the flaring layer, and $\theta$ 
is the angle from the vertical. Then the intensity outside the leading edge $I(x)$ 
is given by 
\begin{equation}
I(x) = \alpha J\frac{d\theta}{dx} + I_0 = \frac{\alpha Jh}{x^2+h^2} + I_0  \; \; \; (x>0)
\end{equation} 
and 
\begin{equation}
I(x=0) = \frac{\alpha J}{h} + I_0 + I_p , 
\end{equation}
where $J$ is the energy of irradiation per unit length and per radian, 
$I_0$ is preflare intensity, and  $\alpha$ is a coefficient. 
$I_p$ is the extra component at the 
leading edge that comes from the initial energy deposition layer. 
We determine the parameters $\alpha J/h$ and $h$ by the least square fitting 
of the intensity profile of the ribbon at 02:28, shown in figure \ref{fig:fig2}. 
Note that $I_p$ is not a free parameter because it can be calculated from 
the observed peak intensity once $\alpha J/h$ and $h$ are determined. 

The solid lines in panel (a) and (b) of figure \ref{fig:fig3} are the result of 
fitting. The best fit parameters are 
$h=124$ km and $\alpha J/h$ = 528 counts for G-band, and 
$h=300$ km and $\alpha J/h$ = 1385 counts for 6302I. 
These values of $h$ are actually an upper limit. 
First of all, $h=124$ km for G-band is comparable to the diffraction limit of 
the telescope. In addition, there may be an effect of long-range scatterred light 
that has not be investigated. Such effects unsharpen the images and therefore 
cause an overestimate of $h$. For 6302I, $h=300$ km is well above the diffraction limit. 
But, as mentioned already, 6302I images suffer significant blurring due to the rapid 
motion of the flare ribbon. The apparent velocity of flare ribbon is roughly 
about 7 km s$^{-1}$. During the 14 sec (time to obtain a 6302I image) the 
flare ribbon move about 100 km, which possibly causes the larger $h$ for 6302I. 
Furthermore, effects of projection (not significant because the flare occurs close 
to the equator) and irradiation from the inner part of the ribbon (may be significant) 
were not taken into account, which also results in an overestimate of $h$. 
Thus the sharpness of the leading edge of the flare ribbon indicates that 
the depth of the diffuse ribbon measured from the irradiation source 
($\sim$ energy deposition height) is of the order of 100 km or even less. 

Such small value of $h$ (i.e., smaller the local scale height $\approx 200$ km) 
indicates that the assumptions (5) and (6) are probably irrelevant approximations, 
because the opacity for the irradiating photons is unlikely to change significantly 
in such a small depth. In order to examine the effect of absorption, we tried to 
fit the intensity profile with an absorption factor, namely 
\begin{equation}
I(x)  = \frac{\alpha Jh}{x^2+h^2}\exp{(-b\sqrt{x^2+h^2})} + I_0  \; \; \; (x>0), 
\end{equation} 
where $b$ is related to the absorption coefficient. The absorption factor reduces 
the intensity far away from the irradiating source, and hence fitting with this factor 
would result in larger $h$. However, the least square fitting yielded   
$b=0$, i.e., no absorption. As a consequent, the best-fit values of $h$ and $J/h$ are 
the same.
More accurate modeling requires proper treatment of radiative transfer and emission 
mechanisms,  but we believe that our simple model gives roughly correct 
depth of the WL emission. 

The actual height of the flare emission in the solar atmosphere is uncertain 
because we do not know the height of the irradiation source. However, 
previous works suggest that the WLF emission is powered by relatively low 
($\lesssim$ 20 keV ) electrons, and hence the energy deposition layer is 
in the upper chromosphere \citep{metcalf03, chending06, fletcher07}. 
On the other hand, if there was sufficient energy in more energetic (say, $\geq$100 keV) 
electrons in this flare, the irradiating source may be deeper. 
We calculated the power of energetic electrons using the hard X-ray data 
from RHESSI and the thick target model \citep{brown71}. 
(The RHESSI data was not available until 02:29UT because the satellite was in the night, 
so we used the data after 02:29UT. )
The electron spectral index was $\sim 4.3$, and the power carried by $\geq$ 20 keV 
electrons was $\sim 3 \times 10^{29}$ erg s$^{-1}$, and that carried by $\geq$ 100 keV 
electrons was $\sim 6 \times 10^{27}$ erg s$^{-1}$. 
In order to compare the energetics, we have to calculate the power radiated 
in the WL emissions. This requires spectral information, which is unknown 
in the present observation. We leave it for future study, and only mention that 
the computed power of electrons is comparable to the X5.3 WLF 
studied by \citet{metcalf03}, whose results indicated that the high energy 
($\gtrsim$ 100 keV) electron was not sufficient to power the WLF.

\subsection{Magnetic Transient}
Transient signal in magnetograms during strong flares have been known for decades 
\citep{patterson81, zirin81, kosovichev01}.  Some of such signals are indeed irreversible 
and hence probably real magnetic field change \citep{wang02, yurchyshyn04}, but those 
which appear only during the impulsive phase are likely to be an artifact due to the distortion 
of the spectral lines \citep{patterson84, qiu03}. The nature of latter is still poorly 
understood because of the lack of the spectroscopic observations. 

\begin{figure}
  \begin{center}
    \FigureFile(60mm,60mm){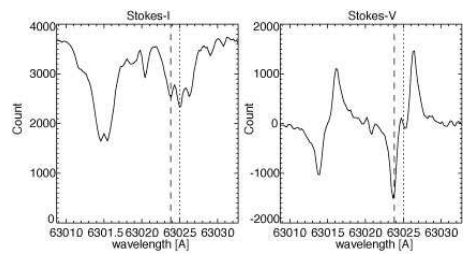}
  \end{center}
  \caption{Stokes I and V profiles of the northern sunspot at 21:02 UT on 2006 December 12 
  obtained by SOT/SP. The vertical lines indicate the line center (dotted) and 
  -120 m\AA \ offset (dashed). 
  }\label{fig:fig7}
\end{figure}

The sign reversal of 6302V count found in our event is co-spatial and co-temporal 
with the flare ribbon in 6302I and hence falls into the latter category. Unfortunately 
there was no SP observation during the flare, but the full Stokes profiles of this 
active region were obtained in the previous day. Figure \ref{fig:fig7} shows 
the Stokes I and V profiles near the location of the flare ribbon in the northern 
sunspot but at 21:02UT on 2006 December 12. The vertical lines indicate 
the line center and -120 m\AA \, offset; the 6302I and 6302V images from FG were 
taken at -120 m\AA. Because of the low temperature in the umbra, the spectral line 
is already highly distorted by molecular lines, which makes it difficult to 
determine the continuum level. A conservative estimate of the continuum level 
is the maximum count in the profile, which is about 3700 in this profile. 

In the leading edge where 6302V showed sign reversal, the 6032I 
count increase about 50\%, from $\sim$ 2500 to $\sim$ 3700. 
Assuming the same line profile as shown in figure \ref{fig:fig7}, the enhanced 
count at -120 m\AA \, is comparable to the pre-flare continuum level. 
We cannot conclude that the line turned to emission because we do not 
know the continuum level during the flare, but it is probably fair to say 
that the line profile was significantly distorted by the flare emission, 
which caused the "magnetic transient" in this event.

\section{Conclusions}
The high-resolution and seeing-free images from the SOT has revealed the very 
fine structures in a WLF (more precisely, G-band and Fe\emissiontype{I} 6302 flare). 
The flare ribbons in G-band and 6302I clearly show two distinct features: 
bright and sharp leading edges (kernels) and following (surrounding) diffuse part. 
In the diffuse part, the pre-flare structure such as umbral dots, penumbral filaments 
and granules remain visible, giving an impression that the diffuse part are 
optically thin. The maximum contrast in the flare ribbons is about 600\% for G-band and 
about 50\% for 6302I. 

Assuming that the leading edge (kernels) is heated directly by nonthermal 
particles beams and the diffuse parts are heating by radiative back-warming, 
we made a simple model to derive the depth of the layer of diffuse emission 
using the intensity profile of the flare ribbon. We found that the layer of 
diffuse emission is about 100 km below the height of the irradiation source. 
If the height of initial energy deposition and hence that of the 
irradiation source is in the upper chromosphere as suggested by 
previous studies \citep{metcalf03, chending06, fletcher07}, 
our result indicates that the height of the diffuse component, 
which we believe due to the radiative back-warming, is also in the 
upper or middle chromosphere. 

The 6302V images (longitudinal magnetogram) show transient sign reversal, 
which is probably the same phenomenon known as magnetic transient. 
The sign reversal was co-spatial and co-temporal with the flare ribbon, 
and probably due to a significant distortion of the Fe I 6302 line during the flare, 
though we cannot be conclusive because of the lack of spectroscopic data. 

\bigskip

Hinode is a Japanese mission developed and launched by ISAS/JAXA, collaborating with NAOJ as a domestic partner, 
NASA and STFC (UK) as international partners. Scientific operation of the Hinode mission is conducted by 
the Hinode science team organized at ISAS/JAXA. This team mainly consists of scientists from institutes in the partner countries. 
Support for the post-launch operation is provided by JAXA and NAOJ (Japan), STFC (U.K.), NASA, ESA, and NSC (Norway).
This work was carried out at the NAOJ Hinode Science Center, which is supported by the Grant-in-Aid for Creative 
Scientific Research ``The Basic Study of Space Weather Prediction" from MEXT, Japan (Head Investigator: K. Shibata), 
generous donations from Sun Microsystems, and NAOJ internal funding. 
H.I. is supported by a Research Fellowship from the Japan Society of the Promotion of Science for 
Young Scientists. He also appreciates the fruitful comments from 
H. Hudson, S. Krucker, T. Yokoyama, and the anonymous referee. 

\end{document}